\documentclass[a4paper,12pt]{article}
\usepackage{amssymb,amsmath}
\usepackage{float}
\pdfoutput=0
\usepackage{amsfonts}
\usepackage{color}
\usepackage{graphicx}

\usepackage{setspace}
\usepackage{authblk}

\newcommand{\pM}{\partial M}
\newcommand{\drm}{\mathrm{d}}

\newcommand{\scalar}[2]{(#1,#2)}
\newcommand{\scalarb}[2]{\langle#1,#2\rangle}
\newcommand{\bth}{\boldsymbol{\theta}}
\renewcommand{\H}{\mathcal{H}}
\newcommand{\D}{\mathcal{D}}
\newcommand{\nsl}{\vec{\gamma}\cdot\vec{n}\,}

\numberwithin{equation}{section}

\title{\bf Edge States: Topological Insulators, Superconductors and QCD Chiral Bags} 

\author[1]{M. Asorey}
\author[2]{A.P. Balachandran}
\author[3]{J.M. P\'{e}rez-Pardo}
\affil[1]{\small Departamento de F{\'{\i}}sica Te\'orica, Universidad de Zaragoza\\
E-50009  Zaragoza, Spain }
\affil[2]{Physics Department, Syracuse University, Syracuse, NY 13244- 1130 and Instituto Internacional de Fisica, Natal, Brasil}
\affil[3]{Departamento de Matem\'{a}ticas, Universidad Carlos III de Madrid, Avda. de la Universidad 30, 28911 Legan\'{e}s, Madrid, Spain.}

\begin{document}

\date{}


\maketitle
\begin{abstract}
The dynamics of   the magnetic field in a superconducting phase is described by an effective  massive bosonic field theory. If 
the superconductor is confined in a domain $M$ with  boundary $\pM$, the  boundary conditions of the electromagnetic fields are predetermined by physics. They are  time-reversal and also parity invariant for adapted geometry. They lead to edge excitations while in comparison, the bulk energies have a large gap. A similar phenomenon occurs for topological insulators where appropriate boundary conditions for the Dirac Hamiltonian also lead to similar edge states and an ``incompressible bulk''. They give spin-momentum locking as well. In addition time-reversal and parity invariance emerge for adapted geometry. Similar edge states appear in QCD bag models
with chiral boundary conditions.
\end{abstract}

\section{Introduction}
The formulation of quantum theories of massless tensor fields on a spatial  Riemannian manifold $M$ requires the self-adjointness and positivity of its negative Laplacian $-\nabla^2$. Both properties are readily fulfilled if $M=\mathbb{R}^N$ or a compact boundary-less Riemannian manifold, and are the basis of massless tensor field theories on Minkowski space.

But there are many situations where the manifold $M$ is not $\mathbb{R}^N$, but is a compact Riemannian manifold with boundary $\pM$. Such manifolds are important for quantum Hall and Casimir effects and topological insulators. 
If $M$ has a boundary $\pM$, the Laplace operator $-\nabla^2$ may fail to satisfy $-\nabla^2\geq 0$. Thus, for a scalar field, let $\vec{n}$ be the outward-drawn unit normal vector and $\dot{\psi}:=\vec{n}\cdot\nabla\Psi|_{\partial M}$ on $\partial M$. Consider the Robin boundary condition 
\begin{equation}\label{BC}
\dot{\psi}=\mu\psi=\mu \Psi|_{\partial M}
\end{equation}
on $\pM$ with $\mu>0$ and where $\psi=\Psi|_{\partial\Omega}$ denotes the boundary value of $\Psi$. Let us also introduce the useful notations
\begin{equation}\label{scalar}
\scalar{\Psi}{\Phi}:=\int_M\drm V_M\Psi(x)^\dagger \Phi(x),
\end{equation}
\begin{equation}\label{scalarb}
\scalarb{\psi}{\phi}:=\int_{\pM}\drm V_{\pM}\psi(x)^\dagger \phi(x)
\end{equation}
where $\drm V_{M}$ and $\drm V_{\pM}$ are respectively the volume forms for the metric $g$ entering the Laplacian operator
 $\nabla^2$, and its pull-back to $\pM$. Then 
\begin{equation}\label{Green}
	\scalar{\Psi}{-\nabla^2\Psi}=\scalar{\nabla \Psi}{\nabla\Psi}-\scalarb{\psi}{\dot{\psi}}=\scalar{\nabla \Psi}{\nabla\Psi}-\mu\scalarb{\psi}{\psi}\;.
\end{equation}
The last term here is negative for positive $\mu>0$.
For this reason, provided that $\mu$ is large enough, there exist $\Psi$ such that $\scalar{\Psi}{-\nabla^2\Psi}<0$ and $-\nabla^2$ has negative eigenvalues \cite{Ba93}. The corresponding eigenstates are edge states localised in a small neighborhood of $\pM$. In particular, Asorey et. al. \cite{As05}, proved that as $\mu$ becomes large, and the Dirichlet condition is approached, the negative eigenvalues recede to $-\infty$. At the same time, the corresponding eigenstates get progressively more localised at the edge and eventually become weakly zero as $\mu\to\infty$. Numerical evidence of this phenomenon in one dimension and a numerical algorithm to solve such eigenvalue problems can be found in \cite{Ib13}.\\

The purpose of this paper is to report on our studies of this remarkable phenomenon which happens for all tensor fields, including the electromagnetic field. It is also insensitive to the topology and dimension of $\pM$. We show in Section 2 that the boundary conditions \eqref{BC} naturally arise when $M$ and $\mathbb{R}^N\backslash{M}$ support different phases, the former being the massive phase of mass $m\simeq \mu$ of the order parameter. In particular $M$ can support a pseudo-Goldstone boson.

Similar edge states exist for the massive Dirac Hamiltonian as well as shown in Section 3. They may  play a role in the physics of topological insulators. In this case, negative eigenvalues are not a problem for quantum field theory. As before, the bulk states have a large gap compared to the edge excitations. There is also a  spin-momentum locking phenomenon. These are among the desired features of topological insulators \cite{Ha10}.

The foregoing considerations do not depend on the topology or the Riemannian metric of $M$.

Let us now make brief remarks on parity $P$ and time-reversal $T$.

As regards, parity, it  is a global diffeomorphism and for this reason must act on $M$ to be even defined. That depends also 
on the nature of the boundary $\partial M$ which may not be a sphere. It must also be an isometry of the metric of $M$ and its boundary
$\partial M$. For simplicity, to enforce these requirements, let us assume that the metric on  $M$ is flat and that $ M$ is a spherical ball ${\cal B}^d$.
For $d=1$, $M$ is the interval $[0, R]$ with $ \partial M=\{0, R\}$, for $d=2$, $M$ is the disk ${\cal B}^2$ with $ \partial M=S^1$ and for
 $d=3$, $M$ is the ball ${\cal B}^3$ with $ \partial M=S^2$. For such a geometry, for tensor fields, $P$ is a symmetry. So is $T$ regardless of the above geometry.
 
 The Dirac Hamiltonian $H$ with a mass term  $m$ requires separate comments. In this case we will also find a Hamiltonian $H_E$ for edge states which controls the APS-like boundary conditions for $H$. For $d=1$, $H_E$ is just a finite-dimensional matrix, we will not comment on it for now. For $d=3$, if $H$ is based on an irreducible representation of $\gamma$-matrices, breaks $P$ and $T$ even on $\mathbb{R}^2$ \cite{redlich} and hence also on ${\cal B}^2$. At the same time, the edge Hamiltonian $H_E$ is $T$-invariant and also $P$-invariant if $M$ is the spherical disk ${\cal B}^2$ as above. 
 But  for $d=3$, i.e. if $M$ is the spherical ball ${\cal B}^3$, $H_E$ breaks $P$ and $T$, being the massive $2+1$ Hamiltonian on
 the sphere $S^2$. As $H_E$  determines the APS boundary condition, $H$ too breaks $P$ and $T$ symmetries. 
  
 We remark that even for the $d=2$ disk, where $H_E$ preserves  $P$ and $T$  whereas $H$ violates both discrete symmetries,
since  the bulk and edge modes can get coupled by interactions, the  $P$ and $T$ symmetries of $H_E$ are likely to be only approximate, protected perhaps by the mass gap of the bulk.

We can of course recover $T$-invariance for $H_E$ or $H$ if appropriate reducible representations 
of $\gamma$-matrices  are used, as done by Altland and Zirnbauer \cite{altlandzimbauer} (for a review, see \cite{stone}).
Also both symmetries can be preserved if  we consider local boundary conditions which generalise bag chiral boundary conditions 
and which are considered in Section 4.
 Section 5, which is the concluding section, discusses spin Hall effect using our model, as also the Majorana condition for spinorial edge excitations. Finally, we formulate
 the edge Hamiltonian for these spinorial edge modes.

 The work of Altland and Zirnbauer has been generalised by Ryu {\it et al} \cite{Ryu}, Le Clair and Bernard \cite{LeClair} and others. As we discuss in the final section, the analysis of our approach along their lines requires further assumptions about the boundary manifold and the symmetries of the dynamics of the boundary theory. The edge effects we find come from the nature of the boundary conditions for the Hamiltonian. While their detailed properties can depend on discrete symmetries as in their work, their existence and the presence of a large gap in the bulk are very general phenomena. They are present not  just for spinorial systems, but also for all tensorial systems.

For reasons of clarity, we focus  only on scalar and spinor fields in this paper, reserving tensor fields and certain mathematical details to a paper in preparation \cite{As13}.

 \section{Scalar Fields}

We assume henceforth that $M$ is compact, with a Riemannian metric $g$ and a smooth boundary of codimension 1.\\

The existence of negative energy edge states of $-\nabla^2$ for the boundary conditions \eqref{BC} has long been known 
\cite{Li63,Ba93,Ba95} and has also been recently studied by \cite{TRG+Ercolessi} in the context of black holes. A general demonstration of their existence is due to \cite{As05} and goes as follows:\\

Introduce Gaussian normal coordinates in a collar neighbourhood $X$ of $\pM$ with $r\in[1-\epsilon,1]$ the radial and $\bth=\{\theta^a\}$ the angular coordinates, $(r=1,\bth)$ being the coordinates of $\pM$. Then on $X$, the metric $g$ and the Laplacian take the form
\begin{equation}\label{metric collar}
	g(r,\bth)= 
		\begin{pmatrix}
			1 & 0 \\ 0& \Omega(r,\bth)
		\end{pmatrix}
\end{equation}
\begin{equation}
	\nabla^2=\partial^2_r+\frac{1}{\sqrt{|\Omega(r,\bth)|}}\partial_a\sqrt{|\Omega(r,\bth)|}\partial_a\equiv \partial^2_r+\nabla^2_{\bth}\;,
\end{equation}
where $\partial_a:=\frac{\partial}{\partial \theta^a}$ and $|\Omega(r,\bth)|=\operatorname{det}\Omega(r,\bth)$. Set $s=\frac{\pi}{2\epsilon}(1-r)$ and 
\begin{equation}\label{Edge}
	\Psi(x)=
		\begin{cases}
			\xi(\bth) \exp(-\frac{2\mu\epsilon}{\pi}\tan s), & (s,\bth)\in X\\
			0, &  (s,\bth)\in M\backslash X
		\end{cases}\;,
\end{equation}
where $\xi$ is any smooth function of $\bth$. This function verifies the boundary condition \eqref{BC}.

Asorey {\it et al} \cite{As05} show that for $\epsilon$ small enough 
\begin{equation}\label{AIMinequality} {
	\scalar{\Psi}{-\nabla^2\Psi}\leq {\pi \over  2\epsilon}\left(  {1\over 4 k } (1+ \delta)-
{k\over 2}\, (1- \delta)\right)  \scalarb{\xi}{\xi}+\epsilon(1+ \delta)\scalarb{\xi}{-\nabla_{\bth}^2\xi}
	}
\end{equation}
and 
\begin{equation}\label{AIMinequality2} {
(\Psi,\Psi) \geq
{\pi(1 - \delta)\over  4\epsilon (k+1)} \, \scalarb{\xi}{\xi}\, 
	}
\end{equation}
where $k=2\epsilon\mu/\pi$ and  $\delta$ comes from the bound of  the variation of the metric, $|\Omega(r,\Gamma)| <
|\Omega(0,\Gamma)|(1 + \delta)$,  in the collar  around the boundary for  small enough $\epsilon$.
Both inequalities provide an upper bound for the energy of the state \eqref{Edge}
\begin{equation}\label{AIMinequality3} {
E= \frac{\scalar{\Psi}{-\nabla^2\Psi}}{(\Psi,\Psi)}\leq {2 (k+1)}\left(  \frac{1+ \delta}{4 k (1-\delta)} -
{k\over 2}\, +\frac{2\epsilon^2}{\pi}  \frac{1+ \delta}{1-\delta}\frac{\scalarb{\xi}{-\nabla_{\bth}^2\xi}}{\scalarb{\xi}{\xi}} \right),
	}
\end{equation}
which shows that the edge states \eqref{Edge}  have negative energies for large enough $\mu$. In fact, $E\to-\infty$ as $\mu\to\infty$ when we approach Dirichlet boundary conditions.

The localisation of $\Psi^\dagger \Psi$ near the boundary gets sharper as $\mu$ gets larger,  and its width can be made as small as we please by choosing a large enough $\mu$. Hence $\Psi$ approaches zero weakly in the Dirichlet limit $\mu\to\infty$. In fact the convergence to a null vector is strong in this limit \cite{As05}.

\subsection*{Interpretation}

If $M$ is a superconductor and $\mathbb{R}^N\backslash M$ is a dielectric or vacuum, the order parameter $\Psi$ in the different
effective theories (London, Ginsburg-Landau or Anderson-Higgs) is obtained as a solution of a second order differential equation
involving the Laplacian of $M$ and an effective mass $m$. The De Gennes boundary conditions  \cite{DeGennes} are in fact Robin boundary conditions \eqref{BC} with $\mu<0$.  The Meissner effect states that the static magnetic potentials $A_i$ decay exponentially in $r$ from $\pM$ as one goes inwards the superconductor: i.e. the photon acquires a mass $m$ on $M$. The boundary 
conditions in this case can be also chosen to be Robin boundary conditions \eqref{BC}, but in this case with $\mu>0$. 
The choice $m\simeq \mu$ matches the connection of the penetration depth with the effective mass of the electromagnetic 
field into the superconductor in $M$ according to Anderson-Higgs effective model.

But then \emph{all} the wave functions in the domain of $-\nabla^2$ fulfill \eqref{BC}. The general theory requires this for the self-adjointness of $-\nabla^2+m^2$. That in turn predicts that there are low-lying edge states at the interface of a superconductor and the vacuum or a dielectric. 

Nevertheless, the bulk states are gapped. 
We can show this as follows. It is enough to consider the scalar $\Psi$. 
Genuine bulk states vanish at the boundary:
\begin{equation}\label{BCbulk0}
	\Psi|_{\pM}=0.
\end{equation}
They can also satisfy the boundary condition \eqref{BC} provided that their normal derivatives 
also vanish at the edge, $	\dot\psi=0$.
Hence 
\begin{equation}
\scalar{\Psi}{(-\nabla^2+m^2)\Psi}=\scalar{\nabla\Psi}{\nabla\Psi}+m^2\scalar{\Psi}{\Psi}\geq m^2\scalar{\Psi}{\Psi}\;,
\end{equation}
where we used \eqref{BCbulk0} during partial integration. What has happened is that the addition of $m^2$ to $-\nabla^2$ lifts the negative energy edge levels of $-\nabla^2$ above zero while at the same time pushing up the positive energy bulk levels above $m^2$.

\emph{It should be possible to check these conclusions experimentally.}

It is interesting that changing $\mu$ to $-\mu$ in \eqref{BC} interchanges the roles of $M$ and $\mathbb{R}^N\backslash M$, with edge states appearing at the boundary of $\mathbb{R}^N\backslash M$ when approached from the side of $\mathbb{R}^N\backslash M$.

Any system with a scalar order parameter $\Psi$ in $M$ supporting a massive phase with broken symmetry and  a massless phase with intact symmetry $\mathbb{R}^N\backslash M$ will have an analogous behaviour. The boundary condition \eqref{BC} guarantees the existence of a  static solution with exponential decay at the boundary, and edge states for the low lying stationary excitations. The crucial requirements for edge excitations are that $M$ has a compact regular boundary and supports a massive field with a static solution. Candidates for this field from the Standard Model would be its pseudo-Goldstone bosons like the pion, but it looks unrealistic to imagine that they can be confined to a manifold with boundary.

Finally, we note that \eqref{BC} is $T$-invariant, $\mu$ being real, and orientation-reversal invariant because $\vec{n}\cdot\nabla$ has that property.

These considerations are very general. They do not depend on the topology and the Riemannian geometry of $M$, requiring only that $\pM$ is a regular codimension 1 boundary. They hold for all tensor fields, in particular for the electromagnetic field. 
Although we have explicitly shown this result only for a scalar field in this paper, it is easy to extend it to $A_i$ as we show in a companion paper.

\subsection*{Example: Disk of Radius $R$}

We next work out the edge states on a disk of radius $R$ with a flat metric. Previous literature has looked at this example, although perhaps only briefly \cite{Ba93,Ba95}.

We thus consider the eigenvalue problem
\begin{subequations}
\begin{equation}
	(-\nabla^2+m^2)\Psi=E^2\Psi,\quad m>0\;,
\end{equation}
\begin{equation}
	\dot{\Psi}|_{\pM}=\mu \Psi|_{\pM},\quad \mu>0\;.
\end{equation}
\end{subequations}
As our focus is on edge states, we shall look for $E^2/m^2\simeq 0$, which implies in particular that $E^2<m^2$. We will also adjust $\mu$ so that $E^2>0$ as we do not want a negative eigenvalue for $-\nabla^2+m^2$. We will see that $\mu^2 \lesssim m^2$ as expected.

In radial coordinates, the eigenvalue problem reads
\begin{subequations}\label{LB disk}
	\begin{equation}\label{LB disk eq}
		\left(-\frac{1}{r}\frac{\partial}{\partial r}r\frac{\partial}{\partial r}-\frac{1}{r^2}\partial_\varphi^2+m^2\right)\Psi=E^2\Psi
	\end{equation}
	\begin{equation}\label{BC disk}
		\partial_r\Psi(R,\varphi)=\mu\Psi(R,\varphi)\;.
	\end{equation}
\end{subequations}
Substituting $\Psi=\mathcal{R}_l(r)e^{i l\varphi}$ and $E=E_l$, one gets
	\begin{equation}\label{radial}
		-\frac{\partial^2}{\partial r^2}\mathcal{R}_l(r)-\frac{1}{r}\frac{\partial}{\partial r}\mathcal{R}_l(r)+\frac{l^2}{r^2}\mathcal{R}_l(r)=(E_l^2-m^2)\mathcal{R}_l(r)\;,
	\end{equation}
	\begin{equation*}
		\dot{\mathcal{R}}_l(R)=\mu \mathcal{R}_l(R)\;.
	\end{equation*}

With 
\begin{equation}
\varepsilon_l^2:=E_l^2-m^2,
\end{equation} 
and $\rho={\varepsilon_l}r$,
 this equation becomes the well-known Bessel equation
	\begin{equation}\label{bessel equation}
		-\rho^2\frac{\partial^2}{\partial \rho^2}{\mathcal{R}}_l(\rho)-\rho\frac{\partial}{\partial\rho}{\mathcal{R}}_l(\rho)+(l^2-\rho^2){\mathcal{R}}_l(\rho)=0\;,
	\end{equation}
with solutions $\mathcal{R}_l(r)=J_l({\varepsilon_l}r)$.

We need to impose now the boundary condition. With the standard notation that {primes} denote differentiation with respect to the argument of the function, we get for the normal derivatives 
\begin{equation}
\frac{\partial \mathcal{R}_l}{\partial r}(r)={\varepsilon_l}J'_l({\varepsilon_l}r)\;,
\end{equation}
and therefore the boundary condition becomes:
	\begin{equation}\label{eigenvalues}
		\mu=\frac{{\varepsilon}_l J'_l({\varepsilon}_lR)}{J_l({\varepsilon}_lR)}.
	\end{equation}
The eigenvalues will correspond to solutions of these equations. If $\varepsilon_l^2>0$ the equation above has solutions near the zeros of the Bessel function in the denominator, and hence an infinite set of them. However here $\varepsilon_l^2$ can be a negative number. Such solutions will correspond to the edge states.

\begin{figure}[h]
 \centerline{  \includegraphics[height=5.5cm]{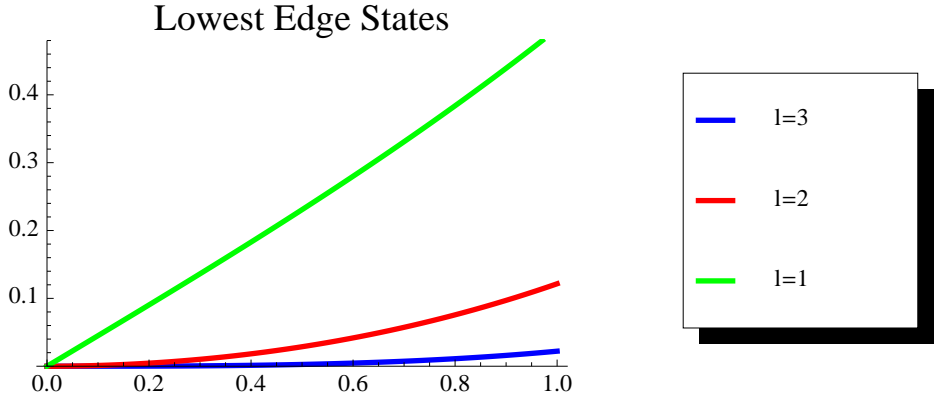}}
 \caption{{Edge states with boundary condition \eqref{eigenvalues} for
 $m=\frac1{R}$ and different values of angular momentum $l=1,2,3$.  
 The horizontal axis represents the dimensionless parameter $\frac{r}{R}$}.}
 \label{fig2}
  \end{figure}

Hence as we focus just on edge states, let us look for solutions with $\varepsilon_l^2=E^2-m^2<0$ or 
\begin{equation}
{\varepsilon_l}=i\lambda_l
\end{equation}
 with $\lambda_l>0$. Then, the condition (\ref{eigenvalues}) can be rewritten in terms of the modified Bessel functions: 
	\begin{equation}
		\mu=\frac{\lambda_l I'_l(\lambda_l R)}{I_l(\lambda_l R)}\;.
	\end{equation}
These equations have solutions for small values of $l$. 
Plots in Figure \ref{fig2}  of the Bessel functions for the corresponding values of $\varepsilon_l^2$ show that 
there are indeed edge states.

\subsection*{Remark}

The scalar products $\scalar{\cdot}{\cdot}$ and $\scalarb{\cdot}{\cdot}$ in \eqref{scalar} and \eqref{scalarb} define two Hilbert spaces $\H(M)$ and $\H(\pM)$, one for $M$ and one for $\pM$. We need them in the next section.

We remark that our considerations are valid if $\Psi$'s and $\Phi$'s are multicomponent, then $\Psi^\dagger\Phi=\sum_\alpha\Psi_\alpha^*\Phi_\alpha$. The operator $\nabla$ can also contain a connection.

\section{The Dirac Hamiltonian}

The Dirac operator on the manifold $M$ of dimension $d$, Riemannian metric $g$ and mass $m$ is the Hamiltonian $H$ of the $(d+1)$-dimensional Dirac operator. It reads
\begin{equation}
	H=-i\gamma^i\nabla_i+m\gamma^{d+1},\quad i=1,\dots,d\;,
\end{equation}
where the gamma matrices verify
\begin{subequations}\label{gamma matrices}
\begin{equation}
	\gamma^i\gamma^j+\gamma^j\gamma^i=2g^{ij}\;,
\end{equation}
\begin{equation}
	\gamma^i\gamma^{d+1}+\gamma^{d+1}\gamma^i=0\;,
\end{equation}
\begin{equation}
	(\gamma^{d+1})^2=\mathbb{I}\;,
\end{equation}
\begin{equation}
	(\gamma^\nu)^\dagger=\gamma^\nu,\quad \nu=1,\dots,d+1\;.
\end{equation}
\end{subequations}
We hereafter set $g^{ij}=\delta^{ij}$ for simplicity so that we get the usual anticommutation relations for the gamma matrices. An important \emph{formal} property of $H$ is that 
\begin{equation}
	H^2=-\nabla^2+m^2\;.
\end{equation}
While for manifolds without boundary, this identity is meaningful, for manifolds with boundary, one must pay attention to the domains of definition of $H$ and $-\nabla^2$. Nevertheless one can use it as a helpful guide to thought.

The boundary conditions for $H$ can be found by considering the following identity:
\begin{equation}\label{Green Dirac}
	\Sigma(\Psi,\Phi):=\scalar{\Psi}{H\Phi}-\scalar{H\Psi}{\Phi}=i\scalarb{\psi}{\vec{\gamma}\cdot\vec{n}\,\phi}\;.
\end{equation}
Here we follow the notation of \eqref{scalar} and \eqref{scalarb} identifying $\Psi^\dagger\Phi$ with $\sum_\alpha\Psi_\alpha^*\Phi_\alpha$, $\alpha$ being the spinor index.

Now let $K$ be any self-adjoint operator on $\H(\pM)$ with no zero eigenvalue and anti-commuting with $\vec{\gamma}\cdot\vec{n}$, i.e.
\begin{subequations}\label{APS properties}
	\begin{equation}
		K^2>0\;,
	\end{equation}
	\begin{equation}\label{APS property}
		\vec{\gamma}\cdot\vec{n}K=-K\vec{\gamma}\cdot\vec{n}\;.
	\end{equation}
\end{subequations}

Then we can split $\H(\pM)$ into the orthogonal direct sum $\H^{+}(\pM)\oplus\H^{-}(\pM)$ where $\H^{\pm}(\pM)$ are spanned by the eigenvectors of $K$ for positive (negative) eigenvalues.
\begin{equation}
K|_{\H^{+}(\pM)}>0,\quad K|_{\H^{-}(\pM)}<0\;.
\end{equation}
Clearly $\scalarb{\Psi^{(+)}}{\Phi^{(-)}}=0$ if $\Psi^{(+)}\in\H^{+}(\pM)$ and $\Phi^{(-)}\in\H^{-}(\pM)$. 

It follows from \eqref{APS property} that 
	\begin{equation}
		\left(\vec{\gamma}\cdot\vec{n}\right)\,\H^{\pm}(\pM)=\H^{\mp}(\pM)\;.
	\end{equation}
Hence if we impose the boundary condition $\Phi|_{\partial M},\Psi|_{\pM}\in \H^{-}(\pM)$ then 
\begin{equation}\label{Dirac symmetric}
	\scalar{\Psi}{H\Phi}-\scalar{H\Psi}{\Phi}=0\;.
\end{equation}
This shows that $H$ is a symmetric operator. One can easily show that $H$ is self-adjoint as well, hence $H^\dagger=H$. Thus the domain $\D_K$, or equivalently the boundary condition for $H$,  depends on $K$:
\begin{equation}\label{APS domain}
	\D_K=\left\{  \Psi \bigr| \Psi_{\pM}\in \H^{-}(\pM)   \right\}\;.
\end{equation}
There are also routine Sobolev conditions on $\Psi$, but we need not elaborate on them.

\subsection{APS Boundary Conditions}
What shall we choose for $K$?
Atiyah, Patodi and Singer \cite{At75} proceed as follows. In $X$ we can write
\begin{align*}
	H&=-i(\vec{\gamma}\cdot\vec{n})\partial_r-\frac{i}{r}\gamma^{\bth}\cdot\nabla_{\bth}+m\gamma^{d+1}\\
	&:=-i(\vec{\gamma}\cdot\vec{n})\partial_r+A'(m)\;.
\end{align*}	
Here $\gamma^{\bth}$ and $\nabla_{\bth}$ are the tangential components of $\gamma^i$ and $\nabla_i$. Then if
\begin{align*}
	A(m)&=A'(m)|_{r=R}\\
	&=-\frac{i}{R}\gamma^{\bth}\cdot\nabla_{\bth}+m\gamma^{d+1}\;,
\end{align*}
then
\begin{equation}A(m)^2\geq m^2>0\;,\end{equation}
\begin{equation}\vec{\gamma}\cdot\vec{n}\, A(m)=-A(m)\, \vec{\gamma}\cdot\vec{n}\;.\end{equation}
Hence $A(m)$ satisfies conditions \eqref{APS properties} and is a candidate for $K$. 

But  so is the following one-parameter  family which we choose for $K$ :
\begin{equation}\label{Disk APS}
	K(\mu)=i\vec{\gamma}\cdot\vec{n}A(\mu)\,,\quad \mu>0\;,
\end{equation}
that, according to the relations \eqref{gamma matrices}, also satisfies \eqref{APS properties}.

In the work of Atiyah,Patodi and Singer, $\mu$ is set equal to $m$. We introduce $\mu$ so that we can 
choose it to have the most appropriate value for a given problem.

 With the introduction of the parameter $\mu$, we can use $K(\mu)$ 
 to define boundary conditions for $H$ even if $m=0$. This could be a useful remark, for example for index theory.
 
 Another  remark may be made. The operator $i \gamma^{d+1} \nsl $
is, up to a constant,  the large $\mu$ limit of $K(\mu)$, 
and is  the type of Hamiltonian entering the discussion of Altland and Zirnbauer \cite{altlandzimbauer}.

The work of Altland and Zirnbauer was generalised by Ryu {\it et al} \cite{Ryu}, LeClair and Bernard \cite{LeClair} and references therein.
Comments relating their papers and our approach are in the last section.

The motivation for this choice will be explained later. We will first work out examples to show that it seems to have good physical properties.

\subsection*{Remark}

There are three dimensionful parameters in the problem, namely $m$, $\mu$ and $R$. We will see that low-lying edge excitations in an interval or a disk require $\mu\simeq m$. We can always tune $\mu$ to get optimal results.

The presence of the two ``masses'' $m$ and $\mu$ are here the spinorial counterparts of $m$ and $\mu$ for $-\nabla^2+m^2$.

\subsection*{Example: The Half Line $(-\infty,0]$}

Consider the Dirac Hamiltonian $H=-i\sigma_1\partial_1+m\sigma_2$ on the half-line $M=(-\infty,0]$. So $\pM=\{0\}$. 
$M$ is not compact, but $\partial M$  is, and that is enough for $M$ to serve as an example.
We see that \begin{equation}K(\mu)=i\mu\sigma_1\sigma_2=-\mu\sigma_3\;.\end{equation} Therefore the boundary condition requires that
\begin{equation}
	\Psi|_{\pM}=\begin{pmatrix} 1 \\ 0 \end{pmatrix}\;,
\end{equation}
setting a possible constant multiplying the spinor equal to 1.

For $x\to-\infty$, we must have $\Psi(x)\to 0$. With that in mind we get the zero energy solution (``bound state'')
\begin{equation}
	\Psi(x)=e^{mx}\begin{pmatrix} 1 \\ 0 \end{pmatrix},
\end{equation}
\begin{equation}
	H\Psi(x)=-ime^{mx}(\sigma_1+i\sigma_2)\begin{pmatrix} 1 \\ 0 \end{pmatrix}=0,
\end{equation}
which is localised near $x=0$. 

Besides this normalisable solution, $H$ has a set of  {generalised}  eigenfunctions. Together, they form a complete spectral set.

In this case, the domain of $H$ is independent of $\mu$ provided it is positive, since $\H^{-}(\pM)$ is the same for all $\mu>0$.

\subsection*{Example: The Disk}

Let us also look at the two-dimensional disk with flat metric and radius $R$. The Dirac Hamiltonian is in this case
\begin{equation}
	H=-i(\sigma_1\partial_1+\sigma_2\partial_2)+m\sigma_3\;.
\end{equation}
In spherical coordinates, we have
\begin{equation}\label{Dirac disk}
H=-i\sigma_r\partial_r-\frac{i}{r}\sigma_\varphi\partial_\varphi+m\sigma_3\;,
\end{equation}
where if $\hat{r}$ and $\hat{\varphi}$ are the radial and angular unit vectors,
\begin{equation}\sigma_r=\hat{r}\cdot\vec{\sigma}\,,\quad\sigma_\varphi=\hat{\varphi}\cdot\vec{\sigma}\;,\end{equation}
where we choose the sign of $\hat{\varphi}$ so that $\sigma_r\sigma_\varphi=i\sigma_3$.

Thus,
\begin{subequations}\label{disk gamma gaussian}
\begin{equation}
	\sigma_r=\begin{pmatrix}0 & e^{-i\varphi} \\ e^{i\varphi} & 0 \end{pmatrix}\;.
\end{equation}
\begin{equation}
	\sigma_\varphi=\begin{pmatrix}0 & -ie^{-i\varphi} \\ ie^{i\varphi} & 0 \end{pmatrix}\;.
\end{equation}
\end{subequations}

The conserved angular momentum for \eqref{Dirac disk} is
\begin{equation}\label{angular momentum}
	J=-i\frac{\partial}{\partial \varphi}+\frac{1}{2}\sigma_3\;.
\end{equation}
Its eigenstates $\psi_j^{(\pm)}$ for eigenvalues $j\in\{\pm 1/2, \pm 3/2,\dots\}$ and orbital momentum $l=j+1/2$ ($l=j-1/2$) are 
\begin{equation}
l=j+1/2:\qquad	\psi_j^{(+)}=e^{i(j+1/2)\varphi}\begin{pmatrix} 0 \\ 1\end{pmatrix}\;,
\end{equation}

\begin{equation}
l=j-1/2:\qquad	\psi_j^{(-)}=e^{i(j-1/2)\varphi}\begin{pmatrix} 1 \\ 0 \end{pmatrix}\;.
\end{equation}
An eigenstate of $H$ for eigenvalue $E_j$ can thus be written as 
\begin{equation}
	\Psi_j(r,\varphi)=\alpha_j(r)\psi_j^{(+)}+\beta_j(r)\psi_j^{(-)}\;.
\end{equation}
Then $H\Psi_j=E_j\Psi_j$ leads to the equations
\begin{subequations}
\begin{equation}\label{separation Dirac}
	-i\alpha'_j-\frac{i}{r}(j+1/2)\alpha_j+m\beta_j=E_j\beta_j\;,
\end{equation}
\begin{equation}
	-i\beta'_j+\frac{i}{r}(j-1/2)\beta_j-m\alpha_j=E_j\alpha_j\;.
\end{equation}
\end{subequations}
This leads to the second order equation
\begin{equation}\label{Bessel Dirac}
\alpha''_j+\frac{1}{r}\alpha_j'-\Bigl[ (m^2-E^2_j)+\frac{(j+1/2)^2}{r^2} \Bigr]\alpha_j=0.
\end{equation}

If we can solve this equation for $\alpha_j$ and $E_j$, we can find $\beta_j$ from \eqref{separation Dirac}. But to solve \eqref{Bessel Dirac}, we need to formulate our generalised APS conditions.

\subsection*{The Operator $K(\mu)$}

The operator $K(\mu)$ can be read off from \eqref{Dirac disk}:
\begin{align*}
	K(\mu)&=i\sigma_r\bigl[ -\frac{i}{R}\sigma_\varphi\partial_\varphi+\mu\sigma_3\bigr]\\
		&=\frac{1}{R}\sigma_3i\partial_\varphi+\mu\sigma_\varphi\;.
\end{align*}
We can solve the eigenvalue problem for $K(\mu)$ using $\psi^{(\pm)}_j$. If $\phi_j$ is an eigenstate of $K(\mu)$, we write 
\begin{equation}
	\phi_j=a_j\psi_j^{(+)}+b_j\psi_j^{(-)}\;.
\end{equation}
Then noticing that 
\begin{equation}
K(\mu)=-\frac{1}{R}\sigma_3 J +\frac{1}{2R}+\mu\sigma_\varphi,
\end{equation}
 write 
\begin{equation}\left(K(\mu)-\frac{1}{2R}\right)\phi_j=\lambda_j\phi_j\;.\end{equation} That leads to 
\begin{equation}
	\begin{pmatrix}
		j/R-\lambda_j & i\mu \\ -i\mu & -j/R-\lambda_j
	\end{pmatrix}
		\begin{pmatrix} a_j \\ b_j \end{pmatrix}=0\;.
\end{equation}
Hence $\lambda_j$  are the roots of \begin{equation}\lambda_j^2-\mu^2-j^2/R^2=0\end{equation} or \begin{equation}\lambda_j=\pm\sqrt{\mu^2+j^2/R^2}\;,\end{equation} where $\sqrt{\mu^2+j^2/R^2}>0$. 

Since
\begin{equation}\frac{1}{2R}-\sqrt{\mu^2+j^2/R^2}\leq\frac{1}{2R}-\sqrt{\mu^2+1/4R^2}< 0\end{equation}
the negative values of $\lambda_j$ correspond to the negative eigenvalues of $K(\mu)$ as well. 

We now have the eigenstates of $K(\mu)$:
\begin{subequations}
\begin{equation}
	\lambda_j=-\sqrt{\mu^2+j^2/R^2}=-|\lambda_j|:
\end{equation}
\begin{equation}
	\phi_{j,-}=c_j\Bigl( \psi_j^{(+)}+\frac{i}{\mu}\left(\frac{j}{R}+|\lambda_j|\right)\psi_j^{(-)} \Bigr),
\end{equation}
\end{subequations}
\begin{subequations}
\begin{equation}
	\lambda_j=+\sqrt{\mu^2+j^2/R^2}=+|\lambda_j|:
\end{equation}
\begin{equation}
	\phi_{j,+}=d_j\Bigl( \psi_j^{(+)}+\frac{i}{\mu}\left(\frac{j}{R}-|\lambda_j|\right)\psi_j^{(-)} \Bigr),
\end{equation}
\end{subequations}
where $c_j, d_j\in \mathbb{C}$. Note the two-fold degeneracy of each eigenvalue $\lambda_j$ of \newline$K(\mu)-1/2 R:\ \phi_{\pm j,\epsilon}$, are degenerate for each of $\varepsilon=+$ and $\varepsilon=-$.

\subsection*{Boundary Conditions}

Let us next focus now on the boundary conditions for $\alpha_j$ and $\beta_j$. 

Taking into account angular momentum conservation as well, the required boundary conditions are 
\begin{align}
	\alpha_j(R)&=c_j\\
	\beta_j(R)&=c_j\frac{i}{\mu}\left(\frac{j}{R}+|\lambda_j|\right)\;.
\end{align}

\subsection*{Solving for Edge States}

We want edge states with $|E_j|/m\ll1$. Hence set $\varepsilon_j^2=m^2-E^2_j$, with $0\leq\varepsilon_j\leq m$.
Then, up to a constant, \begin{equation}\alpha_j(r)=I_{j+1/2}(\varepsilon_j r)\;,\end{equation} where $I_n(x)$ is the modified Bessel function of order $n$. 

Using \eqref{separation Dirac} and the recursion relation 
\begin{equation}\label{rpm}
\frac{\drm}{\drm x}I_{j+1/2}(x)+\frac{j+1/2}{x}I_{j+1/2}=I_{j-1/2}(x)\end{equation} we get
\begin{equation}\label{pm}
\beta_j(r)=\frac{-i \varepsilon_j}{E_j-m} I_{j-1/2}(\varepsilon_jr)\;,\end{equation} where \begin{equation}
\label{sign}
E_j=\pm\sqrt{m^2-\varepsilon^2_j}\;.\end{equation}
Note that $E_j$ can have either sign.

Choosing $c_j$, which is at our disposal, to be $I_{j+1/2}(\varepsilon_j R)$, we find the equation determining $E_j$ for edge-localised states:
\begin{equation}\label{spectral Dirac} {
	\frac{\mu \varepsilon_j}{m-E_j}I_{j-1/2}(\varepsilon_j R)=I_{j+1/2}(\varepsilon_j R)\left(\frac{j}{R}+|\lambda_j|\right)\;.
}
\end{equation}
If there are no real solutions $E_j$ for this equation, that means that there are no edge states. This is the case if $j>0$ and $E_j<0$ in \eqref{sign} or $j<0$ and $E_j>0$ in \eqref{sign} .

But that is not the case if $j>0$ and $E_j>0$ in \eqref{sign} or $j<0$ and $E_j<0$ in \eqref{sign}. We have solved this equation graphically and found solutions $E_j$ with $|E_j/m|\ll1$. We display the graphs    and comment on them in Figures 2-6.

\begin{figure}[h]
 \centerline{  \includegraphics[height=5.5cm]{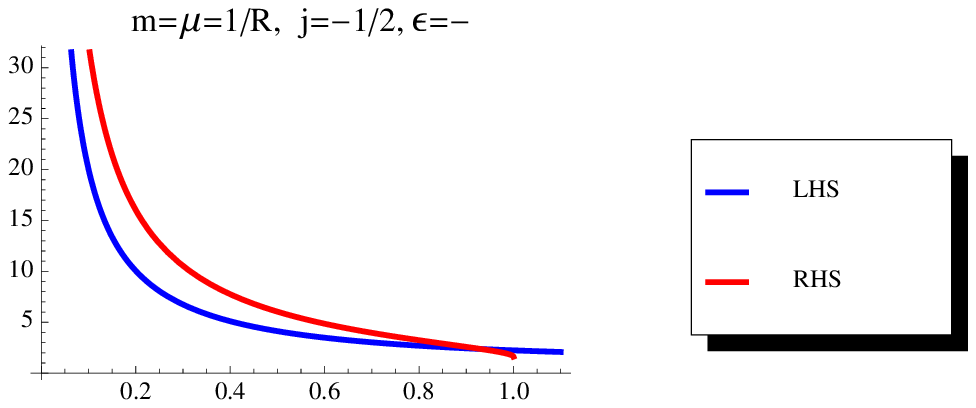}}
 \caption{{ Edge State matching boundary condition \eqref{spectral Dirac} for
 $m=\frac1{R}$ and $j=-\frac12$ with
 negative sign in \eqref{sign}}. 
 The horizontal axis represents the dimensionless parameter $\varepsilon_j R$.  }
  \end{figure}
  \begin{figure}[h]
 \centerline{  \includegraphics[height=5.5cm]{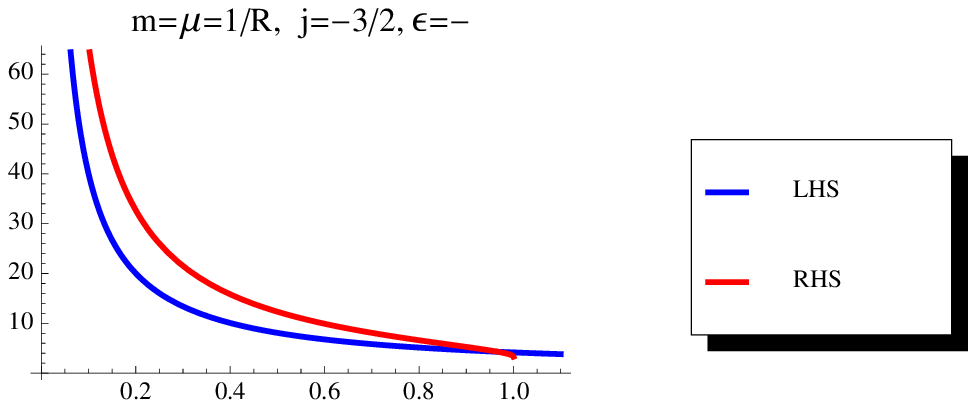}}
 \caption{{ Edge State matching boundary condition \eqref{spectral Dirac} for
 $m=\frac1{R}$ and $j=-\frac{3}{2}$ with
negative sign in \eqref{sign}}. 
The horizontal axis represents the dimensionless parameter $\varepsilon_j R$.  }
  \end{figure} 
\begin{figure}[H]
 \centerline{  \includegraphics[height=5.5cm]{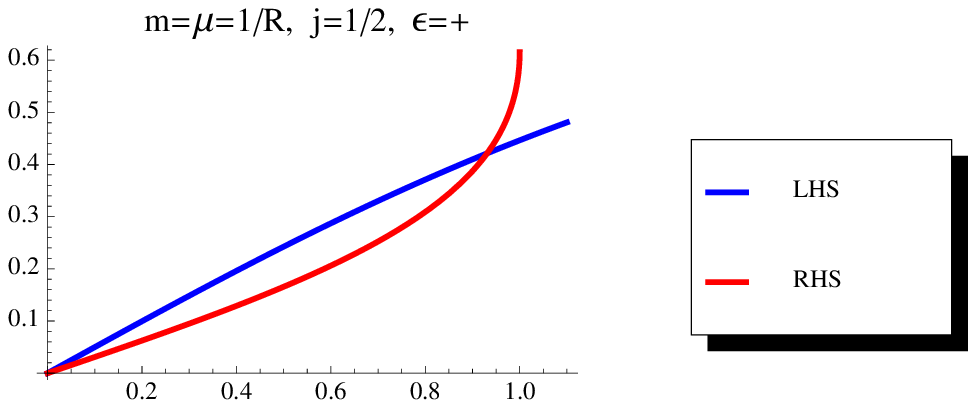}}
 \caption{{ Edge State matching boundary condition \eqref{spectral Dirac} for
 $m=\frac1{R}$ and $j=\frac12$ with
 positive sign in \eqref{sign}}. 
 The horizontal axis represents the dimensionless parameter $\varepsilon_j R$. }
  \end{figure} 
    \begin{figure}[H]
 \centerline{  \includegraphics[height=5.5cm]{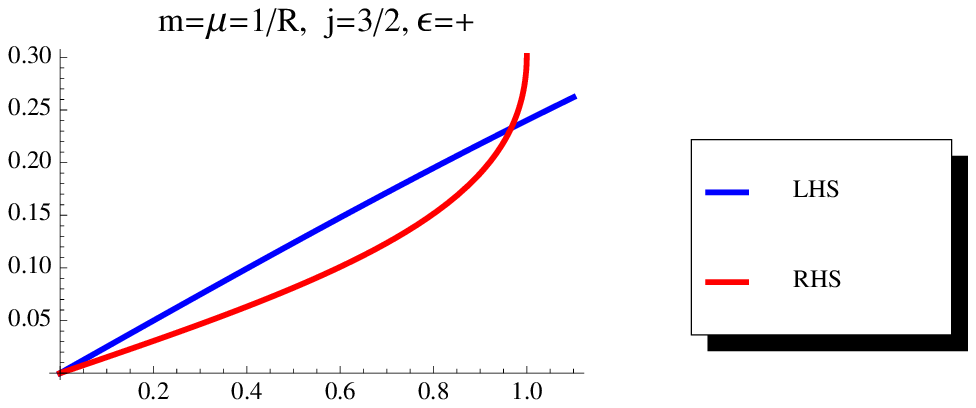}}
 \caption{{  Edge State matching boundary condition  \eqref{spectral Dirac}  for
 $m=\frac1{R}$ and $j=\frac{3}{2}$ with
 positive sign in \eqref{sign}}. 
 The horizontal axis represents the dimensionless parameter $\varepsilon_j R$.}
  \end{figure}
    \begin{figure}[H]
   \includegraphics[height=3.8cm]{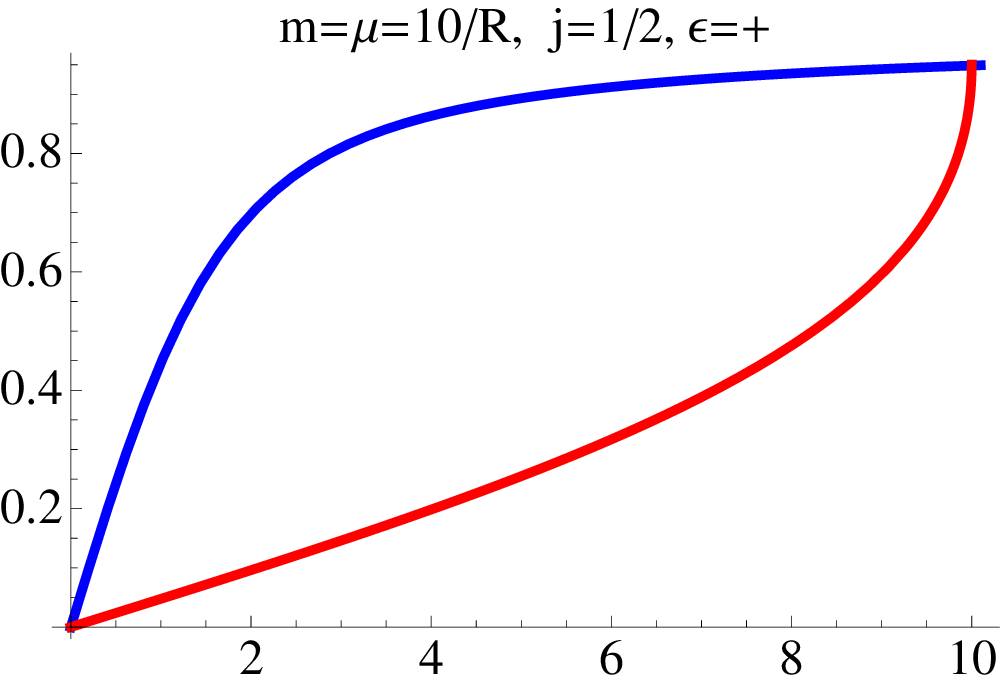}  \hskip1cm   \includegraphics[height=3.8cm]{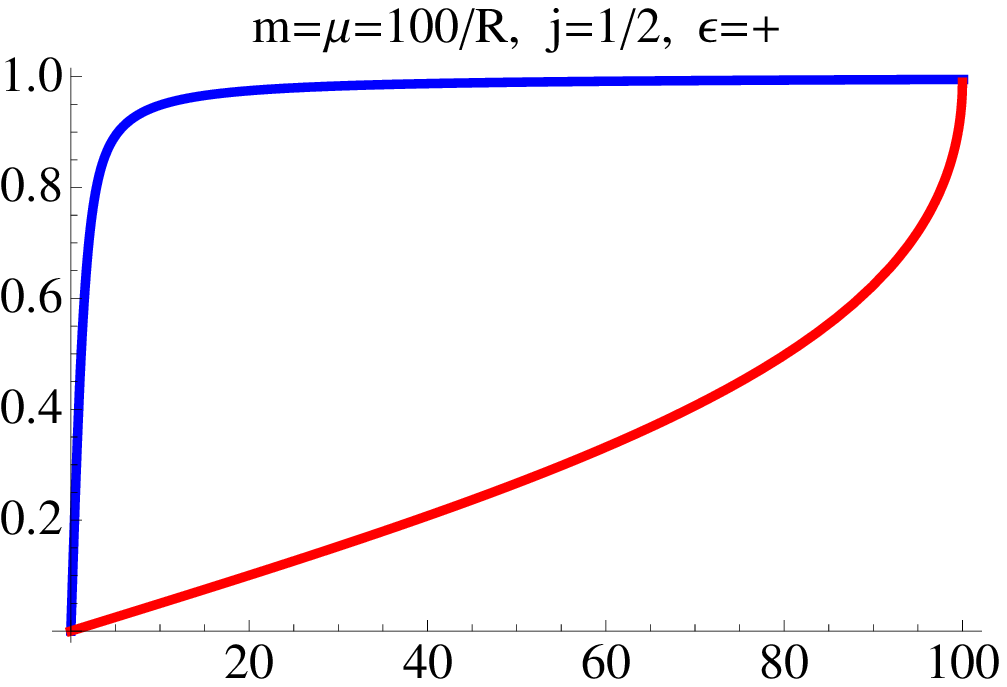}
 \centerline {  \includegraphics[height=3.9cm]{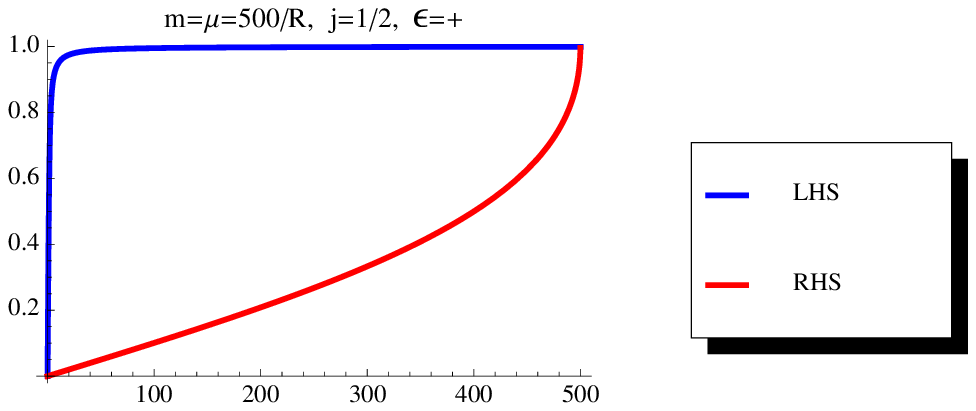}}
 \caption{{  Edge State matching boundary condition  \eqref{spectral Dirac}  for $\mu=m$, $j=\frac{1}{2}$ with
 positive sign in \eqref{sign} and different masses $m=10/R, m=100/R, m=500/R$.}}
  \end{figure}

\subsection*{Spin-Momentum Locking}

An intuitive understanding of the mechanism for spin-momentum locking comes from the expression for $K(\mu)$. In that equation, if \begin{equation}\mu R\gg \langle \sigma_3 J \rangle\;,\end{equation} where $\langle \cdot\rangle$ stands for the mean value in a state, then we expect the $\mu\sigma_\varphi$ term in $K(\mu)$ to dominate. But since \begin{equation}K(\mu)\leq 0\;,\end{equation} by the boundary condition, \begin{equation}\langle \sigma_\varphi \rangle <0\end{equation} on $\pM$. 

Now if $\hat{p}$ denotes the unit vector in the direction of momentum at the boundary, then
\begin{equation}
\langle \sigma\cdot \hat{p}\rangle
	=\begin{cases}
		<0 & \text{if}\quad \hat{p}=\hat{\varphi},\\
		>0 & \text{if}\quad \hat{p}=-\hat{\varphi}
	\end{cases}\;
\end{equation}
by \eqref{disk gamma gaussian}. Thus, if $\hat{p}=\hat{\varphi}$, spin is anti-parallel to $\hat{p}$ and if $\hat{p}=-\hat{\varphi}$, spin is parallel to it.

In the two-dimensional model, this is spin-momentum locking. It leads to net transport of spin anti-parallel to $\hat{\varphi}$ in the direction of $\hat{\varphi}$, see e.g. \cite{Ha10}.

We can quantify this reasoning by computing $\phi_{j,-}^{\dagger}\sigma_{\varphi}\phi_{j,-}$. We get 
\begin{equation}\phi_{j,-}^{\dagger}\sigma_{\varphi}\phi_{j,-}=-\frac{1}{\mu}2|c_j|^2\left(\frac{j}{R}+ |\lambda_j|\right)<0\end{equation} for \emph{all} positive and negative $j$. Hence also 
\begin{equation}\scalarb{\phi_{j,-}}{\sigma_{\varphi}\phi_{j,-}}<0\,,\quad \forall j\;.\end{equation}
This result is stronger than the qualitative reasoning. 

But the mean momenta  in the vector states $\phi_{\pm j,-}$ are antiparallel: 
\begin{equation}\phi_{j,-}^\dagger(-i\partial_\varphi)\phi_{j,-}=|c_j|^2\Bigl[ (j+1/2)+\frac{j-1/2}{\mu^2}\left(j/R +|\lambda_j|\right)^2\Bigr]\;.\end{equation}

This is positive if  \begin{equation}\quad \text{}\quad j\in\{1/2,3/2,\dots\}\end{equation} and negative if
\begin{equation}\quad \text{}\quad j\in\{-1/2,-3/2,\dots\}\;.\end{equation} This is in particular true for $j=\pm1/2$ which gives the lowest edge level. 

The result on spin-momentum locking follows.

\subsection*{Remarks}

We note that \begin{equation}\sigma_{\varphi}\H^{(-)}(\pM)\not\subset\H^{(-)}(\pM)\;,\end{equation} \begin{equation}-i\partial_{\varphi}\H^{(-)}(\pM)\not\subset\H^{(-)}(\pM)\;.\end{equation}
Thus they do not preserve the domain $\D_K$ of $H$ implying that they are anomalous  \cite{Es86,Es02,Ba12}

\subsection*{Majorana Condition}

We next show that $K(\mu)$ is invariant under the anti-unitary involution 
\begin{equation}
I:\psi\to \sigma_1\psi^*, \psi\in\H(\pM).
\end{equation}
 The operator $I$ is not charge conjugation, which interchanges $\H^{(-)}(\pM)$ with $\H^{(+)}(\pM)$. The result is shown by calculating $I\phi_j^{(\pm)}$:
 \begin{equation}I\phi_{j,-}=\frac{\bar{c}_j}{{c}_{-j}}(-i/\mu)\bigl[j/R+|\lambda_j|\bigr]\phi_{-j,-},\end{equation}
\begin{equation}I\phi_{j,+}=\frac{\bar{d}_j}{{d}_{-j}}(-i/\mu)\bigl[j/R-|\lambda_j|\bigr]\phi_{-j,+}\end{equation}
These equations already show that $I$ maps $\H^{\pm}(\pM)$ into $\H^{\pm}(\pM)$ and is compatible with our boundary conditions.

Since $\lambda_j=\lambda_{-j}$, $I$ actually commutes with $K(\mu)$ establishing our claim.

\section{Chiral Bag Boundary Conditions}
It is possible to find 
another family of local boundary conditions, similar to MIT bag boundary conditions used in the analysis of
 quark confinement  \cite{mit} or their generalizations like the chiral bag boundary conditions \cite{chiral}.
 We now explain them using the most general boundary conditions which lead to a self-adjoint Dirac Hamiltonian.

  For arbitrary $\Psi$ and $\Phi$, the boundary term \eqref{Green Dirac} can be written as the difference
  of the two chiral components $\Psi_\pm=\frac12(1\pm\nsl)\Psi$  of spinors  $\Psi=\Psi_+ +\Psi_-$, 
   \begin{equation}
 \Sigma(\Psi,\Phi)=i \scalarb{ \Psi_+}{ \Phi_+}- i\scalarb{\Psi_-}{ \Phi_-}.
   \end{equation}
  The most general boundary condition leading to a self-adjoint Hamiltonian $H$   is given in this
 approach by
 \begin{equation}\label{cdos}
 (1-\nsl)\psi=U \gamma^{d+1} (1+\nsl)\psi,
 \end{equation}
 where $U$ is any unitary operator on the boundary Hilbert space of spinors  commuting with $\nsl$. 
 For simplicity, we assume 'local' boundary conditions where U is a finite dimensional matrix acting only on spinor indices.
Now \eqref{cdos} can be expressed as follows using the Cayley transform:
 \begin{equation}
 \label{dbc}
 \nsl \psi= \frac{1-U \gamma^{d+1}}{I+U\gamma^{d+1}} \psi.
 \end{equation}
 A very simple type of chiral boundary conditions  are given by 
  \begin{equation}\label{ubc}
 U=e^{2 i\arctan e^{\theta}},
  \end{equation}
  which because of the identities
    \begin{eqnarray*}\label{ibc}
\!\!\!\!\frac{1-U \gamma^{d+1}}{I+U\gamma^{d+1}}\!\!&=&\!\!\frac{1+U }{I-U}(1-\gamma^{d+1})+\frac{1-U }{I+U}(1+\gamma^{d+1})\\
\!\!&=&\!\!i\cot(\arctan e^{\theta}) (1-\gamma^{d+1})-i \tan(\arctan e^{\theta}) (1+\gamma^{d+1})\\
\!\! &=&\!\!i e^{-\theta} (1-\gamma^{d+1})-i e^{\theta} (1+\gamma^{d+1})= -i e^{\theta\gamma^{d+1}}\gamma^{d+1},
  \end{eqnarray*}
  corresponds the chiral
 bag boundary conditions \cite{chiral,jenalaplata}
 \begin{equation}
\label{cbc}
	\frac12\left(1-i \gamma^{d+1}e^{-\theta\gamma^{d+1}}\nsl\right)  \psi=0.
\end{equation}
A further reminder  may be made. For $\theta = 0$, the operator $ i \gamma^{d+1} \nsl$ 
is the large $\mu$ limit of $K(\mu)$  up to a constant 
and has been discussed by Altland and Zirnbauer \cite{altlandzimbauer}. For further discussion, please see the final section.

 \subsection*{Example: The Half Line $(-\infty,0]$}
 
 Consider the Dirac Hamiltonian $H=-i\sigma_1\partial_1+m\sigma_2$ on the half-line $M=(-\infty,0]$. So $\pM=\{0\}$ as we had
 for APS boundary conditions. 
 The Dirac Hamiltonian 
\begin{equation}
	H=-i\sigma_1 \partial_1+m \sigma_3\;
\end{equation}
 is self-adjoint when its domain obeys the chiral boundary condition
\begin{equation}
\label{ddbc}
	 \psi(0)=  i {\rm e}^{\theta \sigma_3} \sigma_3 \sigma_1 \psi(0).
 \end{equation}
To find the edge state of $H$ with boundary conditions \eqref{ddbc}, we assume that 
 \begin{equation}\label{cocho}
\Psi(x)=	{\rm e}^{\mu x}	\begin{pmatrix}
			 {\rm e}^{\theta/2}  \\ - i  {\rm e}^{-\theta/2} \\ 
		\end{pmatrix},
\end{equation}
where $\mu>0$ to guarantee the normalisability of $\psi$ as a  bound state.
The eigenvalue equation reduces to
 \begin{equation}
\left(- i \mu\sigma_1 +m \sigma_3	\right)\begin{pmatrix}
			{\rm e}^{\theta/2}  \\ - i  {\rm e}^{-\theta/2} \\ 
		\end{pmatrix}
		= E_{} \begin{pmatrix}
			{\rm e}^{\theta/2}  \\ - i  {\rm e}^{-\theta/2} \\ 
		\end{pmatrix},
\end{equation}
 i.e.
  \begin{equation}
	\begin{matrix}
			(m-E_{})  {\rm e}^{\theta/2} - \mu {\rm e}^{-\theta/2}=0  \\   \mu {\rm e}^{\theta/2} - (m+E_{})  {\rm e}^{-\theta/2}=0  \\ 
		\end{matrix},
\end{equation}
which has a solution if and only if
  \begin{equation}
E=m-\mu e^{-\theta}=\mu e^{\theta}-m.
\end{equation}
This implies in particular that for an edge state to exist, the value of $\mu$ is fixed to be
  \begin{equation}
\mu=\frac{m }{\cosh \theta},
\end{equation}
The value of $E$ is also given by  
  \begin{equation}
  E= {m}\ {\tanh \theta},
\end{equation}
in terms of $m$ and $\theta$. 
The solution $\psi$ corresponds to a bound edge state attached to the boundary wall at $x=0$.
An ansatz like \eqref{cocho}  for $\mu < 0$ is not square integrable and does not 
correspond to a physical state vector.

However, with the alternative chiral 
boundary condition 
 \begin{equation}
\label{cbc2}
	\frac12\left(1+i \gamma^{d+1}e^{-\theta\gamma^{d+1}}\nsl\right)  \psi=0
\end{equation}
there is no edge states, that is, no normalisable eigenvector of $H$ localised near $x=0$.
 
 \subsection*{Example: The  Disk $B_2$}

Let us consider a Dirac electron moving on a disk $B_2$ of radius $R$ as we did for
APS boundary conditions. 
The Dirac Hamiltonian 
\begin{equation}
{
	H=-i\sigma_1 \partial_1-i\sigma_2 \partial_2+m \sigma_3\;}
\end{equation}
 is subject to the boundary conditions \eqref{dbc}
\begin{equation}
	 \Psi(R\cos\varphi,R\sin\varphi)= -i \nsl {\rm e}^{\theta \sigma_3} \sigma_3
		\Psi(R\cos\varphi,R\sin\varphi) 
	\label{rbc}
\end{equation}
with $\nsl=\sigma_1 \cos\varphi+\sigma_2 \sin\varphi$. 

The Hamiltonian is essentially self-adjoint in the space of smooth functions satisfying (\ref{rbc}).
Let us consider the spectrum of  the Hamiltonian $H$. The Hamiltonian is invariant under rotations
with generators given by the total angular momentum ${J_3}={L_3}+\frac12{\sigma_3}$. Let us
consider stationary states of the form
\begin{equation}
\Psi(r ,\varphi)=		\begin{pmatrix}
			 {\rm e}^{i\varphi(j-1/2)}\phi_1(r)  \\ -i {\rm e}^{i\varphi(j+1/2)}\phi_2(r) \\ 
		\end{pmatrix}
\end{equation}
satisfying the boundary conditions (\ref{rbc}) which now read
\begin{equation}\label{rbc2}
			\phi_2(R)={\rm e}^\theta \phi_1(R) ,
\end{equation}
and the eigenvalue equation
\begin{equation}
H \Psi(r ,\varphi)=	E \Psi(r ,\varphi).
\end{equation}
Then $\phi_1$ and $\phi_2$ are solutions of the pair of coupled differential equations,
\begin{equation}\label{eq1}
			(-E+m) {\rm e}^{i\varphi(j-1/2)}\phi_1(r)  + (i\partial_1+\partial_2) i {\rm e}^{i\varphi(j+1/2)}\phi_2(r)=0  
\end{equation}
\begin{equation} \label{eq2}
			i(E+m) {\rm e}^{i\varphi(j+1/2)}\phi_2(r)  - (i\partial_1-\partial_2)  {\rm e}^{i\varphi(j-1/2)}\phi_1(r)=0,
\end{equation}
which can be decoupled into a pair of second order differential equations
\begin{equation}
			\left(\partial_r^2+\frac1{r} \partial_r - \frac{l^2}{r^2}\right)\phi_1(r)= (m^2-E^2)\phi_1(r)  \quad {\rm where}\quad\textstyle l=j-\frac12 
\end{equation}
\begin{equation}
			\left(\partial_r^2+\frac1{r} \partial_r - \frac{{l'}^2}{r^2}\right)\phi_2(r)= (m^2-E^2)\phi_2(r)  \quad {\rm where}\quad\textstyle l'=j+\frac12 ,
\end{equation}
with boundary conditions \eqref{rbc2},
which imply by \eqref{eq1} and \eqref{eq2} the Robin boundary conditions 
\begin{equation}
			\phi'_1(R)=\left((E+m){\rm e}^{\theta}+\textstyle\frac1{R}(j-\frac12)\right) \phi_1(R),
\end{equation}
\begin{equation}
			\phi'_2(R)=\left((m-E){\rm e}^{-\theta}-\textstyle\frac1{R}(j+\frac12)\right) \phi_2(R).
\end{equation}

Let us consider the  states with negative kinetic energies ($T=|E|-m$), i.e. $-m<E<m$. 
The simplest one  correspond to $j=-1/2$ and $l=-1$, $l'=0$, i.e.
solutions of 
 \begin{equation}
			\left(\partial_r^2+\frac1{r} \partial_r - \frac{1}{r^2} \right)\phi_1(r)= (m^2-E^2)\phi_1(r)   
\end{equation}
\begin{equation}
			\left(\partial_r^2+\frac1{r} \partial_r \right)\phi_2(r)= (m^2-E^2)\phi_2(r)  ,
\end{equation}
with boundary conditions \eqref{rbc2}
and  Robin boundary conditions\begin{equation}
\label{bc1}
			\phi'_1(R)=\left((E+m){\rm e}^{\theta}-\textstyle\frac1{R}\right) \phi_1(R),
\end{equation}
\begin{equation}
\label{bc2}
			\phi'_2(R)=(m-E){\rm e}^{-\theta} \phi_2(R),
\end{equation}
respectively. Notice that in the limit $\theta\to \infty$, (\ref{bc1}) leads to Dirichlet boundary conditions for $\phi_1$, i.e. $\phi_1(R)=0$ whereas 
(\ref{bc2}) leads to Neumann  boundary conditions for $\phi_2$, i.e. $\phi'_2(R)=0$.
The solutions are the modified Bessel functions of the first kind,
\begin{equation}
\phi_1(r)= I_1\left(r \sqrt{m^2-E^2}\right),		\quad	\phi_2(r)= I_0\left(r \sqrt{m^2-E^2}\right) 
\end{equation}
and correspond to edge states, localised at the boundaries. The dependence of the energy on the boundary condition is derived from the 
boundary conditions (\ref{bc1}) and (\ref{bc2}) and the recursion relation \eqref{rpm},
\begin{figure}[h]
 \centerline{  \includegraphics[height=6.5cm]{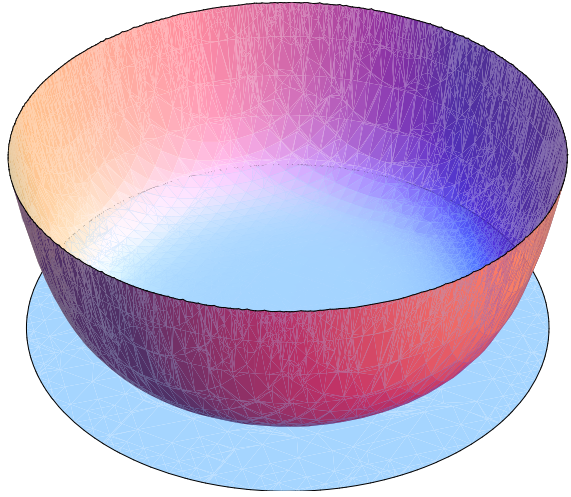}}
 \caption{{Charge density distribution $\Psi^\dagger \Psi$ of edge state with zero energy on a 2D disk $B_2$ for $\theta=0$ and $mR=10$.}}
  \end{figure}
\begin{equation}
\label{bc11}
			\frac{\phi'_1(R)}{ \phi_1(R)}=-\frac1{R}+ \sqrt{m^2-E^2}\frac{I_0\left(R \sqrt{m^2-E^2}\right) }{I_1\left(R \sqrt{m^2-E^2}\right) }=(E+m){\rm e}^{\theta}-\frac1{R},
\end{equation}
\begin{equation}
\label{bc22}
			\frac{\phi'_2(R)}{ \phi_2(R)}= \sqrt{m^2-E^2}\frac{I_1\left(R \sqrt{m^2-E^2}\right) }{I_0\left(R \sqrt{m^2-E^2}\right) }=(m-E){\rm e}^{-\theta},
\end{equation}
which lead to
\begin{equation}
\label{spectrum}
			\sqrt{\frac{m+E}{m-E}}\frac{I_1\left(R \sqrt{m^2-E^2}\right) }{I_0\left(R \sqrt{m^2-E^2}\right) }= {\rm e}^{-\theta}
\end{equation}
In particular, we have a zero mode for $\theta = \log I_0(R m) -  \log I_1(R m) $ which corresponds to  the maximally localised edge state.
The concentration of the  state  on the edge increases as the mass gap increases, which provides the perfect situation for a topological insulator.

  One can find more states with higher angular momenta and negative kinetic energy. They are of the form
  \begin{equation}
		\Psi_n(r ,\varphi)=		 {\rm e}^{i n \varphi }\begin{pmatrix}
			I_n\left(r \sqrt{m^2-E^2}\right) \\- i {\rm e}^{i \varphi} I_{n+1}\left(r \sqrt{m^2-E^2}\right) \\ 
		\end{pmatrix}
\end{equation}
 with $n$ integer. The corresponding energies are given by
  \begin{equation}
			\sqrt{\frac{m+E}{m-E}}\frac{I_{n+1}\left(R \sqrt{m^2-E^2}\right) }{I_n\left(R \sqrt{m^2-E^2}\right) }= {\rm e}^{-\theta}.
\end{equation}
  The number of such states is always finite  and depends on the mass gap. For larger masses, there is a larger number of
  edge states.  For instance for unit mass $m=\frac1{R}$, the number of edge states for $\theta=1,2,3,4$ is $2,7,18,50$, respectively.

Remark: There are similar spinorial edge states for QCD in a three-dimensional ball with chiral boundary conditions. 
The MIT bag model uses the chiral bag boundary conditions in the limit $\theta\to \infty$. In that case there is an infinity of edge states.
In particular the lowest energy state is an edge state. The states of pions and protons  made of  quarks 
localised at the edges of the bag is not  a very  realistic picture for high energies  where according to asymptotic freedom,  
quarks will move freely inside hadron.


\section{Final Remarks}

We plan to do detailed calculations for a spherical three-dimensional ball, where we also hope to investigate the Majorana reality conditions.

Here we next show that there is spin-momentum locking for $d=3$ as well, at least for large $\mu R$. For $d>3$, we expect a result of a similar sort. The reasoning for this expectation, as we already saw for a disk, is that the generalised APS condition does not involve momentum. Also, at least for large $(\mu R)$, the mass term in $K(\mu)$ dominates leading to a momentum independent state vector. The component of spin $\vec{S}$ in a fixed direction $\hat{\varphi}$ tangent to $\pM$ then has a momentum-independent mean value. From this we can infer net spin transfer along $\hat{\varphi}$ for large $\mu$, and later check the exact result as we did on a disk. Unfortunately for $d>3$ we do not know how to extract a vector like $\vec{S}$ from angular momentum generators which are anti-symmetric tensors.

In the final subsection here, we propose a Hamiltonian $H_E$ for the edge excitations. It clearly emerges from our preceding discussions and fulfills the $P$ and $T$ properties mentioned in the Introduction.

\subsection*{Spin-Momentum Locking: The Spherical Ball $B^3$}

As we have postponed the analysis of the eigenvalue problem for $K(\mu)$ to a future date, we focus on the mass term $i\mu\vec{\gamma}\cdot\vec{n}\gamma^{d+1}$ in $K(\mu)$ which dominates for large $\mu R.$

We first do a local analysis. At any point $p\in\partial B^3=S^2$, we can choose $\vec{\gamma}\cdot\vec{n}=\gamma_3$ and $\gamma^{d+1}=\gamma_4$. Then, $i\gamma_3\gamma_4$ is a generator of the $\mathfrak{so}(4)$, or rather the $\mathfrak{spin}(4)$ Lie algebra. The Lie algebra $\mathfrak{so}(4)$ is the direct sum $\mathfrak{su}(2)^{(1)}\oplus\mathfrak{su}(2)^{(2)}$, where $\mathfrak{su}(2)^{(j)}$ are commuting $\mathfrak{su}(2)$ Lie algebras. 

The angular momentum generators $\mathcal{J}_i^{(j)}$ of $\mathfrak{su}(2)^{(j)}$ are
\begin{equation}
	\mathcal{J}_i^{(1)}=\begin{pmatrix}  \tau_i/2 & 0 \\ 0 & 0 \end{pmatrix}\;,
\end{equation}
\begin{equation}
	\mathcal{J}_i^{(2)}=\begin{pmatrix}  0 & 0 \\ 0 & \tau_i/2 \end{pmatrix}\;.
\end{equation}

The generators for the conventional spin $\vec{S}$ are $\vec{\mathcal{J}}^{(1)}+\vec{\mathcal{J}}^{(2)}$:
\begin{equation}
	S_i=\mathcal{J}_i^{(1)}+\mathcal{J}_i^{(2)}=\begin{pmatrix}  \tau_i/2 & 0 \\ 0 & \tau_i/2 \end{pmatrix}\;,
\end{equation}

The generators $M_{i4}$, $i=1,2,3$ of rotations in the $i-4$ plane are $\mathcal{J}_i^{(1)}-\mathcal{J}_i^{(2)}$:
\begin{equation}M_{i4}=\mathcal{J}_i^{(1)}-\mathcal{J}_i^{(2)}=\begin{pmatrix}  \tau_i/2 & 0 \\ 0 & -\tau_i/2 \end{pmatrix}\;.\end{equation}
We can identify $M_{i4}$ with $\frac{1}{2i}\gamma_i\gamma_4$ by setting 
\begin{equation}\gamma_\mu=\begin{pmatrix} 0 & \Sigma_\mu \\ \tilde{\Sigma}_\mu & 0  \end{pmatrix}=(\gamma_\mu)^\dagger\;,\end{equation}
where 
\begin{equation}
\Sigma_i=i\tau_i, \Sigma_4=\mathbb{I}, \tilde{\Sigma}_i=-i\tau_i, \ \tilde{\Sigma}_4=\mathbb{I}. 
\end{equation}

It follows that \begin{equation}K(\mu)\simeq -2\mu (\mathcal{J}^{(1)}_3-\mathcal{J}^{(2)}_3)\;.\end{equation}

The spectrum of $\mathcal{J}^{(j)}_3$ in the four -dimensional Dirac spinor representation is $\{\pm 1/2,0\}$. The boundary condition requires that $K(\mu)<0$, that occurs for the following spinors:
\begin{equation}
	\xi^{(1)}=\begin{pmatrix}1 \\ 0 \\ 0 \\ 0 \end{pmatrix}\;\quad K(\mu)\xi^{(1)}=-\mu\xi^{(1)}\;,
\end{equation}
\begin{equation}
	\xi^{(2)}=\begin{pmatrix}0 \\ 0 \\ 0 \\ 1 \end{pmatrix}\;\quad K(\mu)\xi^{(2)}=-\mu\xi^{(2)}\;.
\end{equation}

The mean values of $\vec{S}\cdot\vec{m}$ for a fixed $\vec{m}$ for the vectors $\xi^{i}$ do not involve momentum, indicating spin transport along $\vec{m}$. 

We also give here the analogues of the above spinors for all $\vec{n}$. Their construction involves the introduction of the Hopf bundle over $\partial B^3= S^2$ \cite{Ba88,Ba91}. Let 
\begin{equation}\zeta=(\zeta_1,\zeta_2)\;,\quad \zeta_i\in\mathbb{C}\;,\end{equation} 
\begin{equation}\label{normalization}
	\sum|\zeta_i|^2=1\;.
\end{equation}

Then set \begin{equation}n_i=\zeta^\dagger\sigma_i\zeta\;.\end{equation} 
\noindent
The normalisation \eqref{normalization} gives $\vec{n}\cdot\vec{n}=1$ as required. The choice $\zeta=(1,0)$ gives $n_3=1$, $n_1=n_2=0$ and leads to our previous considerations. 

Similar results hold for 	
\begin{equation}i\sigma_2\bar\zeta=(\bar\zeta_2,-\bar\zeta_1).\end{equation} 

Now globally we can choose our spinors as follows:
\begin{equation}
	\xi^{(1)}(\zeta)=\begin{pmatrix}\zeta_1 \\ \zeta_2 \\ 0 \\ 0 \end{pmatrix}\;\quad K(\mu)\xi^{(1)}(\zeta)=-\mu\xi^{(1)}(\zeta)\;,
\end{equation}
\begin{equation}
	\xi^{(2)}(\zeta)=\begin{pmatrix}  0 \\ 0 \\ \bar\zeta_2 \\ -\bar\zeta_1 \end{pmatrix}\;\quad K(\mu)\xi^{(2)}(\zeta)=-\mu\xi^{(2)}(\zeta)\;.
\end{equation}

\subsection*{What leads to our Domain $\D_K$?}

For completeness, we also now indicate our considerations leading to the Dirac domain $\D_{K(\mu)}$. 

Let us note that any $K$ we can choose by \eqref{APS properties} acts only on spinors $\xi$ on $\pM$ and not on functions of radial variables. Then if 
\begin{equation}\H(\pM)=\H^{(-)}(\pM)\oplus\H^{(+)}(\pM)\,,\quad \xi \in\H^{(-)}(\pM)\;,\end{equation}
where \begin{equation}K|_{\H^{(+)}(\pM)}>0\;,\quad K|_{\H^{(-)}(\pM)}<0\;,\end{equation} then a wave function $\alpha$  which in $X$ looks like 
\begin{equation}\label{Edge Robin Dirac}
\alpha(x)=\rho(r)\xi(\varphi)
\end{equation}
is a vector in the domain $\D_K$. 

But the Robin boundary condition involves only the radial variable. So we can choose 
\begin{equation}\label{Robin Dirac}
	\dot{\rho}(R)=\nu\rho(R)\;,
\end{equation}
where we distinguish $\nu$ here from the $\mu$ in $K(\mu)$. We can in fact choose $\rho$ to be the function 
\begin{equation}
\rho(r)=e^{-\frac{2\nu\epsilon}{\pi}\tan((1-r)\pi/2\epsilon)}
\end{equation} 
of \eqref{Edge}.  Then \eqref{Edge Robin Dirac} is edge localised and gives low-lying edge states for
$-\nabla^2+m^2$, with $m^2\simeq \nu^2$. Thus it is a good candidate for the edge localised Dirac wave function. 

But, if in fact \eqref{Edge Robin Dirac}  is a good choice, the quadratic form $\scalar{H\Psi}{H\Psi}$ will be small compared to $m^2\scalar{\Psi}{\Psi}$:
\begin{equation}\scalar{H\Psi}{H\Psi}\ll m^2\scalar{\Psi}{\Psi}\;.\end{equation}
Using Green's formula for the Dirac operator and \eqref{Robin Dirac}, we have the identity
\begin{align}
\scalar{H\Psi}{H\Psi}&=\scalarb{-i\vec{\gamma}\cdot\vec{n}\Psi}{H\Psi}+\scalar{\Psi}{H^2\Psi}\\
	&=\nu\scalarb{\Psi}{\Psi}+\scalarb{\Psi}{K(m)\Psi}+\scalar{\Psi}{(-\nabla^2+m^2)\Psi}\label{ffour}\;,
\end{align}
where we have used the fact that \begin{equation}H\Psi|_{r=R}=[-i\vec{\gamma}\cdot\vec{n}\ \nu+A(m)]\Psi,\end{equation}
by \eqref{Robin Dirac}.
The last term on the R.H.S of the second equation is small compared to $m^2\scalar{\Psi}{\Psi}$, but the first term can be large since $\nu^2\simeq m^2$. So we are led to the boundary condition $\Psi\in\D_{K(m)}$ which makes the second term on the R.H.S negative and may become larger than the first term. 

For greater flexibility, we chose instead $\Psi\in\D_{K(\mu)}$ since \begin{equation}\scalarb{\Psi}{K(m)\Psi}=\scalarb{\Psi}{K(\mu)\Psi}+ (m-\mu)\scalarb{\Psi}{i\vec{\gamma}\cdot\vec{n}\gamma^{d+1}\Psi}\end{equation} and the last term is small if $\mu\simeq m$. Thus we finally settle on the domain $\D_{K(\mu)}$. 

{We emphasise that the actual calculations in the earlier sections did not appeal to the arguments here at all.}

{The Hamiltonian $H_E$ for edge excitations} depends on $\mu$, so let us write it as  $H_E(\mu)$. It can be read off from \eqref{ffour}:
\begin{equation}
\label{form}
H_E(\mu)=-K(\mu)-m.
\end{equation}
Here we use $\nu\approx m$. Also for $K(\mu)$, all positive energy levels are projected out, that is they are filled, while it is more conventional to have a filled  
negative energy sea. We have hence judicially  flipped signs in (\ref{form}).

The Hamiltonian $H_E(\mu)$ has an  associated Lagrangian density. For the disk case, it is 
\begin{equation}
\label{fm}
{\cal{L}}_E(\mu)=-\bar\Psi\left(i\sigma_\varphi\partial_t+\frac{i}{R}\sigma_r\partial_\varphi+\mu+m\sigma_\varphi\right)\Psi,
\end{equation}
where $\bar\Psi\:=\Psi^\dagger\sigma_\varphi$,
as is easily shown.

\subsection*{Remarks on earlier work}
 As commented already, in the limit of large $\mu$, our edge Dirac Hamiltonian reduces 
 to a product of $\gamma$ matrices and can be analysed following Altland and Zirnbauer \cite{altlandzimbauer}.
  
  Recent work on topological insulators has been performed by Ryu {\it et al} \cite{Ryu}, Le Clair and Bernard \cite{LeClair} and references therein. Following their lead, we can  examine the symmetry properties  of our edge Hamiltonian as a function of $d$ and $\mu$. We cannot set $m=0$ as that would eliminate the gap in the bulk.
  
  The Dirac operator considered by Le Clair and Bernard  \cite{LeClair} differs from ours because of the presence of the
$m$ term. Its origin is the Robin boundary condition which is central to our work. Their analysis thus needs to be redone in our case.
  
  As regards Ryu {\it et al}, their work seems to require that $\partial M$ is $\mathbb{R}^{d-1}$ and that the potential of the edge Hamiltonian has either the symmetry under  the group $\mathbb{Z}^{d-1}$  or at least it is continuously deformable to a potential with such a symmetry. If that is the case, the momentum space governing the edge Hamiltonian is diffeomorphic to a torus. The occupied energy levels then determine a particular Grassmannian bundle on this torus. Their properties under discrete symmetries are then studied in detail by these authors.
  Our work is not focused on such symmetry properties. Rather   our focus is on  edge states on manifolds with boundaries regardless of symmetries, although we have made occasional comments on symmetries.
    
   It seems possible to extend our work in the directions investigated in the above papers. We plan to explore this possibility.
  

\section*{Acknowledgements}
A.P.\ Balachandran has had extensive discussions on the subject matter of this paper with T.R. Govindarajan, Kumar Gupta, Se\c{c}kin K\"urk\c{c}\"u{o}\v{g}lu, V. P. Nair and  Amilcar Queiroz. The possibility of introducing the parameter $\mu$ in \eqref{Disk APS}  emerged during discussions with Amilcar who also generously helped us with the preparation of the figures. We also thank the JHEP referee for drawing our attention to refs. \cite{Ryu} and  \cite{LeClair}.
APB  was supported by the Institute of Mathematical Sciences, Chennai when this work was in progress. The paper was completed during his stay at the Instituto Internacional de Fisica, Natal, Brasil where he thanks Alvaro Ferraz for hospitality and support.
M. Asorey thanks A. Santagata for discussions. His work has been partially supported by the Spanish MICINN grants FPA2009-09638 and CPAN Consolider Project CDS2007-42 and DGA-FSE (grant 2011-E24/2). J.M. P\'erez-Pardo has been partially
supported by the spanish MICINN grant MTM2010-21186-C02-02 and the
QUITEMAD programme P2009 ESP-1594.

\end{document}